\documentstyle[12pt]{article}
\newlength{\bredde}
\def\slash#1{\settowidth{\bredde}{$#1$}\ifmmode\,\raisebox{.15ex}{/}
\hspace*{-\bredde} #1\else$\,\raisebox{.15ex}{/}\hspace*{-\bredde} #1$\fi}
\newcommand{\beq}{\begin{equation}}
\newcommand{\eeq}{\end{equation}}
\newcommand{\bea}{\begin{eqnarray}}
\newcommand{\eea}{\end{eqnarray}}

\def\gtwid{\raise.3ex\hbox{$>$\kern-.75em\lower1ex\hbox{$\sim$}}}
\def\ltwid{\raise.3ex\hbox{$<$\kern-.75em\lower1ex\hbox{$\sim$}}}
\begin{document}
\topmargin -1.8cm
\oddsidemargin -0.8cm
\evensidemargin -0.8cm
\title{\Large{
Qualitons from QCD}}

\vspace{0.5cm}

\author{{\sc P.H. Damgaard} \\
CERN -- Geneva \\
and \\
{\sc R. Sollacher} \\
Gesellschaft f\"{u}r Schwerionenforschung GSI mbH \\
P.O. Box 110552, D-64220 Darmstadt, Germany
}
\maketitle
\vfill
\begin{abstract} Qualitons, topological excitations with the quantum
numbers of quarks, may provide an accurate description of what is
meant by constituent quarks in QCD. Their existence hinges crucially
on an effective Lagrangian description of QCD in which a pseudoscalar
colour-octet of fields enters as a new variable. We show here how such
new fields may be extracted from the fundamental QCD Lagrangian using
the gauge-symmetric collective field technique.
\end{abstract}
\vfill
\vspace{5.5cm}
\begin{flushleft}
CERN--TH-7073/93 \\
November 1993 \\
hep-ph/9311244
\end{flushleft}
\newpage


A few years ago Kaplan \cite{Kaplan}\footnote{See also the recent work
\cite{Gomelski}.} introduced the notion of ``qualitons",
quark solitons that can arise as topological excitations of a chiral
Lagrangian in the space of (for one flavour) $SU(3)$ {\em colour}.
They may have all the correct quantum numbers of ordinary quarks. Far
from being of academic interest, such coloured Skyrmions hold the
promise of providing a meaningful field-theoretic definition of
what is meant by constituent quarks in the context of strong interactions.

The problem of constituent quarks lies in the almost bizarre success
of the nonrelativistic quark model. Requiring light-quark masses on the
order of $\sim$ 350 MeV, it seems as if the constituent $u$ and $d$
quarks have very little to do with the $u$ and $d$ {\em current} quarks
of the fundamental QCD Lagrangian. The discrepancy is already much
softened in the case of the $s$ quark, and from the $c$ quark and beyond
it seems to a very large extent justified to treat the strong interactions
between $q\bar{q}$ pairs in the nonrelativistic approximation. Indeed,
at that level of mass, the distinction between current quarks and
constituent quarks has almost vanished. To a certain extent this
only sharpens the point that the huge difference between current and
constituent quarks of the lightest quark flavours is very poorly understood.
The chiral quark model \cite{Manohar} is one hybrid that provides a
bridge between the fundamental QCD Lagrangian and the nonrelativistic
quark model. Various aspects of the constituent quark degrees of freedom
and their relation to QCD are also discussed in ref. \cite{several}.

When generalized to $N_f$ flavours (and $N_c$ colours), Kaplan's
proposal amounts to finding the coloured Skyrmions of an
$SU(N_f\times N_c)_L\times SU(N_f\times N_c)_R$ chiral Lagrangian
coupled to the remaining gluonic degrees of freedom. It is
remarkable that in the bosonizable case of 2-dimensional QCD,
the $U(N_f\times N_c)$ bosonization scheme provides
an explicit realization of Kaplan's idea \cite{Ellis}: constituent
quarks can there be understood as the topologically non-trivial
solitons of the bosonized field. The abelian analogue of this
is the more well-known boson-fermion duality in 2 dimensions.

The fact that the qualiton idea emerges naturally in the essentially
solvable case of QCD$_2$ is strong motivation for seeking
a similar construction in 4-dimensional
QCD. Kaplan's original proposal required the introduction of an
effective Lagrangian describing the colour orientation of the
condensate of $\bar{q}q$. This was partly motivated by the
observation that at zero gauge coupling the QCD Lagrangian (of,
for simplicity, one flavour) is invariant under the curious symmetry
of ``global chiral colour" $SU(N_c)_L\times SU(N_c)_R$. Assuming
that this symmetry is spontaneously broken to diagonal
$SU(N_c)$ by a chiral condensate $\langle\bar{q}q\rangle$, it is
natural, in the absence of strong gauge couplings, to describe the
colour chiral dynamics by an effective chiral Lagrangian. Thus, by
introducing a unitary matrix $\Sigma(x)$ of the form
\beq
\Sigma(x) = \exp[2i\Pi^a(x)T_a/F] ~,
\eeq
where $\Pi^a(x)$ is an octet of pseudoscalar fields,
one is led to postulate an effective colour chiral Lagrangian of the
usual $SU(N_f)_L\times SU(N_f)_R/SU(N_f)$ kind, with essentially
only the number of flavours $N_f$ substituted by the number of
colours $N_c$. Such an effective Lagrangian is an expansion in
the derivatives of $\Sigma$, $i.e$., valid in the low momentum limit.
One very crucial difference between the case of spontaneously
broken flavour symmetry and this broken chiral colour is, however,
to be found in the Wess-Zumino term, which embodies the anomaly
content of the underlying Lagrangian. Since there is only one
colour for each quark in the anomaly diagram, the coefficient of
the Wess-Zumino term in the effective Lagrangian differs from the
usual case \cite{Witten} by a factor of $1/N_c$. Being directly related
to the baryon number of the Skyrmion, this modified coefficient
of the Wess-Zumino term now gives rise to a
topological excitation with the correct quantum numbers of the
quarks themselves, instead of the $N_c$-quark baryon. It is this
topological excitation that Kaplan identifies with a constituent
quark.

Perhaps the most serious objection that can be raised against this
picture is the lack of a direct connection with the fundamental QCD
Lagrangian. The colour chiral symmetry is obviously severely broken
by the gluonic interactions, and it is precisely these strong
interactions which are responsible for the formation of the chiral
condensate. A delicate trade-off thus seems to be required, in which
the colour chiral symmetry is significant enough to yield important
information about the ground state of the theory, and yet so strongly
broken by explicit terms in the Lagrangian that a chiral condensate
can be formed at all. In his first paper Kaplan \cite{Kaplan}
approached this problem from the point of view that one can imagine
separating the scales of colour confinement and chiral symmetry
breaking at will by introducing new interactions in the QCD Lagrangian
of the Nambu--Jona-Lasinio type. One can then subsequently tune these
new artifical couplings to zero, while tracing the fate of the
excitations of the theory. But the dynamics of the colour orientation
of the chiral condensate remains to be postulated without direct
connection to the underlying QCD Lagrangian.

The purpose of the present note is to sketch how this connection
between qualitons and QCD may be established. As we shall see, the
existence of an octet of ``coloured pions" in the space of fields
describing QCD dynamics is, surprisingly,
not entirely unnatural. Using a collective field
technique based on field-enlarging transformations and an associated
induced gauge symmetry \cite{Alfaro}, we shall show how one can derive
explicitly the couplings of these new fields to QCD degrees of freedom.
This technique, the ``gauge-symmetric approach to effective Lagrangians"
\cite{us} is based on an {\em exact} rewriting of the fundamental QCD
Lagrangian with ultraviolet cut-off $\Lambda$.\footnote{In two dimensions
the same technique can be used to derive the rules of ``smooth
bosonization"\cite{us1}. Smooth bosonizations allows one to perform
2-dimensional bosonization
or fermionization to any degree one wants, in contrast to the usual
boson-fermion equivalence which is limited to either purely bosonic or
fermionic descriptions.} It is most remarkable
that the resulting effective Lagrangian contains precisely
the kind of terms required to support the coloured Skyrmion
idea of Kaplan. However, our effective Lagrangian does also contain a
full set of chiral-colour rotated quark fields which still interact with
the gluons {\em and} now also with the (for $N_c = 3$)
colour-octet of pseudoscalars
suggested by Kaplan. To make full contact with Kaplan's work, and to
show that these coloured Skyrmions may be stable objects at
short-to-intermediate distances,
requires that the left-over quark degrees of freedom be
integrated out of the path integral. This one way or another assumes
the existence of an
approximation scheme suitable for the problem, and we shall not enter into
details of such a calculation in this paper.

Our starting point is a generating functional of the full QCD
Lagrangian in the chiral limit, $i.e.$ with zero current quark masses,
\beq
{\cal Z}_\Lambda[V,A] ~=~ \int\! d\mu[B][d\bar{q}][dq]\;
\exp\left[-\int\! d^4x\; {\cal L}[V,A]\right]
\label{eq:Zgen}
\eeq
and
\beq
{\cal L}[V,A] = \frac{1}{2g^2} Tr G_{\mu\nu}G_{\mu\nu} +
\bar{q}(\slash{\partial} - i\slash{B} - i\slash{V} - i\slash{A}\gamma_5)q~.
\label{eq:LQCD}
\eeq
Here, $B_\mu = B_\mu^a T_a$ is the gluonic vector field with $T_a$
being the generators of $SU(N_c)$ with $[T_a ,T_b ]=if_{ab}^c T_c$ ,
normalized by $Tr T_a T_b = \frac{1}{2}\delta_{ab}$. The corresponding
field strength tensor is denoted by
\beq
G_{\mu\nu} = \partial_{[\mu} B_{\nu ]} -i [B_\mu , B_\nu ]~~,
\eeq
and $V_\mu(x)$ and $A_\mu(x)$ are vector and axial vector sources of the
corresponding quark currents.
We have explicitly indicated that the generating functional is defined for
the theory with an ultraviolet cut-off $\Lambda$. The choice of this
regulator is not arbitrary; we demand that it be consistent
(see, $e.g.$, ref. \cite{Ball} for a discussion of this point).
To be precise, we shall employ a set of Pauli-Villars regulator fields
associated with each of the quark fields. This suffices to regularize
the quark dynamics to at least one loop, which is all we need for the
present paper. The gluon dynamics must be regularized similarly, but the
details of that regulator will play no r\^{o}le in the following. The
functional measure $d\mu[B]$ of the gluon potential $B_\mu(x)$ includes
the integration over Yang-Mills ghost fields and their exponentiated
Faddeev-Popov determinant. Again, such details of the gluon dynamics
have no direct bearing on the details of the effective Lagrangian we shall
derive in this paper, and there is no need to display these standard terms.

We start with the case of just one flavour, and
now perform a local colour-chiral transformation on the quark fields. It
is most convenient to do this in the slightly asymmetric way of, $e.g.$, a
purely left-handed transformation (with $q_L \equiv P_+q,~ \bar{q}_L
\equiv \bar{q}P_-$ and $P_{\pm} = \frac{1}{2}(1 \pm \gamma_5)$),\footnote{
After having completed this paper, we learned that Frishman, Hanany and
Karliner \cite{Frishman} have extracted the leading $\Sigma$-dynamics from
QCD along similar lines.}
\beq
q_L(x) = e^{2i\Pi(x)/F}\chi_L(x) ~,~~~~~\bar{q}_L = \bar{\chi}_L(x)
e^{-2i\Pi(x)/F} ~.
\label{eq:trans}
\eeq
Here $\Pi(x) = \Pi^a(x)T_a$. We shall for convenience consider
transformations in $U(N_c)$ instead of $SU(N_c)$; the reason for this
will become clear shortly.
The transformation (5)
has two consequences: one is at the classical level of the QCD Lagrangian,
the other in the functional measure of the quark fields. The
classical modification of the Lagrangian is found trivially, and the quantum
mechanical change due to the fermionic measure can be extracted from the
literature (see, in particular, ref. \cite{Ball}, and references therein).
This latter modification of the Lagrangian can be organized as an expansion
in inverse powers of the ultraviolet cut-off $\Lambda$, each term being
computable in a systematic manner.

For the external sources and the gluon potentials, it is
convenient to use the combinations
\beq
L_\mu = B_\mu + V_\mu + A_\mu ~,~~~ R_\mu = B_\mu + V_\mu - A_\mu~,
\eeq
instead of $V_\mu, B_\mu$ and $A_\mu$.

After the colour-chiral rotation, the generating functional takes the form
\begin{eqnarray}
{\cal Z}_\Lambda [V,A] &=& \int\! {\cal D}_\Lambda
[\bar{\chi},\chi ] d\mu [B]\; e^{-\int\! d^4x\; {\cal L}'} \cr
{\cal L}' &=& \bar{\chi} \gamma_\mu (\partial_\mu -iL_\mu^\Pi P_+ -i
R_\mu P_- ) \chi + {\cal L}_J + {\cal L}_{WZ} + {\cal L}_{YM}
\label{eq:ZnewL}
\end{eqnarray}
One ``classical'' effect of this chiral rotation is a modification of
$L_\mu$ of the form
\beq
L_\mu^\Pi = L_\mu  + i\Sigma^\dagger {\cal D}_\mu \Sigma
\eeq
with ${\cal D}_\mu = \partial_\mu -i [L_\mu ,\cdot]$. The other effect of
this chiral rotation is the appearance of the additional terms ${\cal
L}_J$ and ${\cal L}_{WZ}$ in the Lagrangian; they are induced by the
fermionic measure and therefore of purely quantum mechanical origin.

The ``positive parity'' part of this addtional terms can be ordered
as an expansion in inverse powers of the ultraviolet cut-off
$\Lambda$:
\beq
{\cal L}_J = \Lambda^2 {\cal L}_2 + {\cal L}_0 +
\frac{1}{\Lambda^2} {\cal L}_{-2} +
\frac{1}{\Lambda^4}{\cal{L}}_{-4} + \ldots
\eeq

The precise form of the higher order terms is not of interest here,
but can in any case readily be calculated using the technique
described in ref. \cite{Ball}. Let us here list only the first two terms,
\bea
{\cal L}_2 &=& \frac{\kappa_2}{4\pi^2} tr\; {\cal A}_\mu^{(s)}
{\cal A}_\mu^{(s)} \mid_{s=1}^0 \cr
{\cal L}_0 &=&
\frac{1}{8\pi^2} tr\; \biggl( -iF_{\mu\nu}^{(s)}
[{\cal A}_\mu^{(s)} ,{\cal A}_\nu^{(s)} ] + \frac{1}{3} {\cal D}_\mu^{(s)}
{\cal A}_\nu^{(s)} {\cal D}_\mu^{(s)} {\cal A}_\nu^{(s)} - \frac{2}{3}
({\cal A}_\mu^{(s)} {\cal A}_\mu^{(s)} )^2 \cr &&\qquad + \frac{4}{3}
{\cal A}_\mu^{(s)} {\cal A}_\nu^{(s)} {\cal A}_\mu^{(s)} {\cal
A}_\nu^{(s)} \biggr) \mid_{s=1}^0~~,
\label{eq:J+L}
\eea
where $tr$ denotes the trace over colour indices (for the case of one
flavour). The covariant derivative ${\cal D}_\mu$ has already been
defined. The other fields in these expressions are defined by
\bea
{\cal A}_\mu &=& \frac{1}{2} ( L_\mu - R_\mu )\cr
F_{\mu\nu} &=& \frac{1}{2} (\partial_{[\mu} L_{\nu ]} -i [L_\mu ,L_\nu
] + \partial_{[\mu} R_{\nu ]} -i [R_\mu ,R_\nu ]) ~~.
\eea
The auxiliary parameter $s$ of the chiral transformation appears only
in connection with $L_\mu$:
\beq
L_\mu^{(s)} = L_\mu + i e^{-2is\Pi/F} {\cal D}_\mu e^{2is\Pi/F}
\eeq
Finally, the coefficient $\kappa_2$ in eq.
(\ref{eq:J+L}) is a regularization-scheme dependent
constant. In general, for the Pauli-Villars scheme adhered to in this
paper, this is only one out of a series of coefficients $\kappa_n$ given by
\beq
\kappa_n = \sum_i c_i k_i^n \ln k_i^2 \quad , \quad \sum_i c_i = 1 \quad ,
\quad \sum_i c_i k^m_i = 0 \quad \hbox{for}\quad m = 1,\ldots ,4
\eeq
where the $k_i$'s are the Pauli-Villars regulator masses in
units of the cutoff $\Lambda$, $i.e.$ $M_i = k_i\Lambda$.

The leading term of the ``negative parity'' part is the
integrated Bardeen-anomaly, the Wess-Zumino term:
\bea
{\cal L}_{WZ} &=& \frac{i}{16\pi^2 F} \int_{1}^{0} ds
\;\epsilon_{\mu\nu\rho\sigma}\; tr\; \Pi \biggl(
F_{\mu\nu}^{(s)} F_{\rho\sigma}^{(s)} + \frac{1}{3}
 A_{\mu\nu}^{(s)}  A_{\rho\sigma}^{(s)} \cr &&\qquad +
\frac{8i}{3} ( F_{\mu\nu}^{(s)} {\cal A}_\rho^{(s)} {\cal A}_\sigma^{(s)}
+ {\cal A}_\mu^{(s)} F_{\nu\rho}^{(s)} {\cal A}_\sigma^{(s)} + {\cal
A}_\mu^{(s)} {\cal A}_\nu^{(s)} F_{\rho\sigma}^{(s)} )
\cr && \qquad + \frac{32}{3} {\cal A}_\mu^{(s)} {\cal A}_\nu^{(s)}
{\cal A}_\rho^{(s)} {\cal A}_\sigma^{(s)} \biggr) + {\cal O} (\Lambda^{-2})
\label{eq:J-L}
\eea
Here, $A_{\mu\nu}$ is defined as
\beq
A_{\mu\nu} = {\cal D}_{[\mu} A_{\nu]} = \frac{1}{2} (\partial_{[\mu}
L_{\nu ]} -i [L_\mu ,L_\nu ] + \partial_{[\mu} R_{\nu ]} -i [R_\mu
,R_\nu ]) ~~.
\eeq

It is convenient to express this chirally rotated QCD
Lagrangian in terms of the $\Sigma$-field as far as possible. The
expressions above can then be written in the following manner:
\bea
{\cal L}_2 &=& -\frac{\kappa_2}{16\pi^2} tr\! D_\mu \Sigma^\dagger
D_\mu \Sigma +
\frac{i\kappa_2}{8\pi^2} tr\; (L_\mu^{(ext)} \Sigma D_\mu
\Sigma^\dagger + R_\mu^{(ext)} \Sigma^\dagger D_\mu \Sigma ) + \ldots
\cr
{\cal L}_0 &=& -\frac{1}{32\pi^2} tr\! \biggl[ -i G_{\mu\nu} (D_\mu
\Sigma^\dagger D_\nu \Sigma + D_\mu \Sigma D_\nu \Sigma^\dagger )
+ \frac{1}{3} D_\mu D_\nu \Sigma^\dagger D_\mu D_\nu \Sigma \cr &&+
\frac{1}{6} D_\mu \Sigma^\dagger D _\nu \Sigma D_\mu \Sigma^\dagger
D_\nu \Sigma - \frac{1}{3} D_\mu \Sigma D_\mu \Sigma^\dagger D _\nu
\Sigma D_\nu \Sigma^\dagger \biggr] + \ldots
\label{eq:LSig+}
\eea
Here we have used the abbreviation $
D_\mu = \partial_\mu -i[B_\mu,~\cdot~]$.
The dots in (\ref{eq:LSig+}) denote terms of higher order in the
external sources as those
which have been displayed. The external sources $L_\mu^{(ext)}$ and
$R_\mu^{(ext)}$ are defined as $V_\mu \pm A_\mu$ analogous to $L_\mu$
and $R_\mu$.

Similarly to the ``positive parity'' part, the ``negative parity''
Wess-Zumino term can be expressed in a colour-gauge covariant manner
as (compare also \cite{Pak})
\bea
{\cal L}_{WZ} &=& -\frac{i}{48\pi^2} \int_0^1\! ds\;
\epsilon_{\mu\nu\rho\sigma}\; tr \biggl[ \frac{2\Pi}{F} D_\mu
\Sigma^{(s)\dagger} D_\nu \Sigma^{(s)}
D_\rho \Sigma^{(s)\dagger} D_\sigma \Sigma^{(s)} \cr
&& - i G_{\mu\nu} \Bigl( D_\rho \Sigma^{(s)\dagger} D_\sigma
\Sigma^{(s)} \frac{\Pi}{F} - D_\rho \Sigma^{(s)\dagger} \frac{\Pi}{F}
D_\sigma \Sigma^{(s)} + \frac{\Pi}{F} D_\rho \Sigma^{(s)\dagger} D_\sigma
\Sigma^{(s)} \cr
&& + D_\rho \Sigma^{(s)} D_\sigma \Sigma^{(s)\dagger}
\frac{\Pi}{F} - D_\rho \Sigma^{(s)} \frac{\Pi}{F} D_\sigma
\Sigma^{(s)\dagger} + \frac{\Pi}{F} D_\rho \Sigma^{(s)} D_\sigma
\Sigma^{(s)\dagger} \Bigr) \cr
&& - 2 \frac{\Pi}{F} G_{\mu\nu} G_{\rho\sigma} - \frac{1}{2}
G_{\mu\nu} \Bigl( \Sigma^{(s)\dagger} G_{\rho\sigma} \Sigma^{(s)} +
\Sigma^{(s)} G_{\rho\sigma} \Sigma^{(s)\dagger} \Bigr) \biggr]\cr
&& -\frac{i}{48\pi^2} \epsilon_{\mu\nu\rho\sigma}\; tr
\biggl[ R_\mu^{(ext)} \Bigl( D_\nu \Sigma^{\dagger} D_\rho
\Sigma D_\sigma \Sigma^{\dagger} \Sigma
+ i G_{\nu\rho} \Sigma^{\dagger} D_\sigma \Sigma +
i \Sigma^{\dagger} D_\nu \Sigma G_{\rho\sigma}
\cr &&+ \frac{i}{2} \Sigma^{\dagger} G_{\nu\rho} D_\sigma
\Sigma - \frac{i}{2} D_\nu \Sigma^{\dagger} G_{\rho\sigma}
\Sigma \Bigr) -L_\mu^{(ext)} \Bigl( D_\nu \Sigma D_\rho
\Sigma^{\dagger} D_\sigma \Sigma \Sigma^{\dagger} \cr &&
+ i G_{\nu\rho} \Sigma D_\sigma \Sigma^{\dagger} +
i \Sigma D_\nu \Sigma^{\dagger} G_{\rho\sigma}
 + \frac{i}{2} \Sigma G_{\nu\rho} D_\sigma
\Sigma^{\dagger} - \frac{i}{2} D_\nu \Sigma G_{\rho\sigma}
\Sigma^{\dagger} \Bigr)\biggr] + \ldots
\label{LSig-}
\eea
The terms omitted are either of
second order in the external sources or of order $1/\Lambda^2$, at
least according to the naive counting of powers of the cut-off.\footnote{
In the quantum theory, the naive counting of inverse powers of the
ultraviolet cut-off $\Lambda$ may be modified by compensating
cut-off dependent matrix elements.}

An important observation can now be made concerning baryon or quark
number. For that purpose consider the current coupled to an external
vector source $V_\mu$, now taken to be abelian, before and
after the transformation (\ref{eq:trans}). We find
\beq
\bar{q} \gamma_\mu q = \bar{\chi} \gamma_\mu \chi - \frac{i}{24\pi^2}
\epsilon_{\mu\nu\rho\sigma} \; tr\; \Bigl( D_\nu \Sigma D_\rho
\Sigma^\dagger D_\sigma \Sigma \Sigma^\dagger +\frac{3i}{2} G_{\nu\rho}
( \Sigma D_\sigma \Sigma^\dagger - \Sigma^\dagger D_\sigma \Sigma
)\Bigr) + \ldots
\label{eq:shiftVC}
\eeq
where the omitted terms are again formally of order $1/\Lambda^2$. This
implies that the baryon number
\beq
B_q = \frac{1}{N_c}\int\! d^3x \; q^\dagger q = B_\chi + \frac{1}{N_c}
\frac{i}{24\pi^2} \epsilon_{ijk}
\; tr\; \int\! d^3x \;  \partial_i \Sigma^\dagger \partial_j
\Sigma \partial_k \Sigma^\dagger \Sigma ~~,
\label{eq:shiftB}
\eeq
$i.e.$, that the original baryon number
carried by $\bar{q},q$ is now the sum of the baryon number of
$\bar{\chi}, \chi$ and $1/N_c$ of the topological charge of $\Sigma$. For the
moment the latter is completely arbitrary and so is the quark number
of $\bar{\chi}, \chi$. However, if we insist on integer topological
charges for $\Sigma$ then it is obvious that we can shift the quark number
of $\bar{\chi},\chi$ by integers; topological charge represents
quark number.

So far all we have done is to  make a local chiral rotation of the
quark fields in colour space. The phases $\Pi^a(x)$ are as yet
completely decoupled from QCD dynamics. To turn these new fields into
genuine collective fields of the QCD generating functional, we
integrate over $\Sigma(x)$ in the path integral, using the usual left
and right invariant Haar measure. Since the generating functional is
independent of $\Sigma(x)$, this is a valid procedure that does not
change any physics. It does, however, imply a huge redundancy in the
path integral: Since for each slice of $\Sigma(x)$ the path integral
is {\em independent} of $\Sigma(x)$, we are integrating over a continuum of
identical copies. This is the Lagrangian manifestation of a hidden
local gauge symmetry which is uncovered in the collective field
approach \cite{Alfaro}. In terms of our variables here, the local
gauge symmetry reads as follows:
\begin{eqnarray}
\chi_L (x) &\to& e^{2i\alpha (x)} \chi_L (x)\cr
\bar{\chi}_L (x) &\to& \bar{\chi}_L (x) e^{-2i\alpha (x)} \cr
\Sigma(x) &\to& \Sigma(x) e^{-2i\alpha (x)}
\label{eq:symL}
\end{eqnarray}
where $\alpha (x)$ is a local transformation parameter
in the same representation of $SU(N_c)$ as $\Pi(x)$. In principle, we
can allow these gauge transformations to be topologically nontrivial,
thus shuffling quark number from $\bar{\chi}, \chi$ to $\Sigma$ and
vice versa.

We can make constructive use of this local
symmetry by means of imaginative gauge-fixings. This is the essential
principle behind the gauge-symmetric approach to effective Lagrangians
\cite{us,us1}. When we gauge-fix on $\Pi^a(x) = 0$, we simply recover
QCD in terms of the original formulation. But there is an immense
freedom to choose more interesting gauges; each of these will correspond
to QCD in a partially field-redefined version. If one wishes to
entertain the idea of baryon number being carried by topologically
non-trivial $\Sigma$-configurations, the objective must be to find a
gauge in which baryon number of the colour-chirally rotated quark
fields $\chi$ and $\bar{\chi}$ is zero. Does such a gauge exist?

Consider a gauge in which $\chi^\dagger\chi = 0$. Naively, such
a gauge would be invalid since it would restrict the {\em physical}
baryon number density $q^\dagger q$ to vanish as well. However, because
of the induced
Wess-Zumino term we get a new contribution to the latter. From eq.
(\ref{eq:shiftB}) it follows that
\beq
q^\dagger q = - \frac{i}{24\pi^2} \epsilon_{ijk} \; tr\;
\Bigl( D_i \Sigma D_j \Sigma^\dagger D_k \Sigma
\Sigma^\dagger +\frac{3i}{2} G_{ij} ( \Sigma D_k
\Sigma^\dagger - \Sigma^\dagger D_k \Sigma )\Bigr) + \ldots
\label{eq:constr}
\eeq
in a gauge where $\chi^\dagger\chi = 0$. Baryon number is thus
entirely carried by the $\Sigma$-fields. The terms omitted are of
higher order in $1/\Lambda^2$.

How many gauge-fixing conditions do we need? Strictly speaking, none.
Since we are integrating $\Sigma(x)$ over the Haar measure of $SU(N_c)$,
the path integral is well-defined despite the local gauge symmetry.
If we do not fix {\em any} degrees of freedom, we have, however, no
means of attaching physical significance to the collective fields
$\Sigma(x)$. If we again consider a quantity like baryon number, it
can still be shared between the $\chi$-fields and the $\Sigma$-fields,
but the fraction carried by each cannot be given any physical
interpretation. Only the {\em sum} (\ref{eq:shiftVC}) is colour-chiral gauge
invariant, and only this sum enters into physical matrix elements.
Thus, if we do not gauge-fix at all, we will have no possibility of
identifying specific $\Sigma$-field configurations (such as coloured
Skyrmions) with physical excitations.\footnote{The reader may worry
that this seems to imply that specific gauges can be given definite
physical meaning, contrary to the feeling that physics
should not depend on the gauge chosen. This is one of the crucial
aspects of the gauge-symmetric approach to effective Lagrangians
\cite{us}: Although final answers do {\em not} depend on the chosen
gauge (or, equivalently, the chosen field basis for the theory),
the physical picture realizing these final answers may be strongly
dependent on the gauge. The purpose is precisely to find a field
parametrization as close as possible to the physical excitations
of the theory, independently of the original basis of fields.}
Of course, if we simply integrate out these new degrees of freedom
$\Sigma$ from the functional integral, we always, by construction,
recover QCD in the original formulation.
Having introduced $N_c^2$ collective fields $\Pi^a(x)$, we are
nevertheless free
to introduce up to $N_c^2$ gauge-fixing conditions in the path
integral. If we just implement {\em one}, we only constrain a certain
combination of $\Sigma$-fields, while all gauge rotations that keep
this combination invariant are still permitted. This may not be the
optimal situation, especially if one wishes to perform approximate
calculations, and it can also easily lead to ill-defined ghost terms in
the action, due to residual gauge symmetries in the ghost sector.
Let us therefore instead
consider in some detail the consequences of imposing a complete gauge
fixing of all $N_c^2$ collective degrees of freedom. We choose it to
include the particular gauge fixing $\chi^\dagger\chi = 0$, as suggested by
our earlier arguments. Both of these requirements can be fulfilled by
the gauge in which the zeroth component of the coloured currents
$J^a_0(x) = \bar{q}(x)T^a\gamma_0 q(x)$ of
the {\em original} quark fields are all described in terms of collective
fields only. For such a gauge to be valid, it is clear from eq.
(\ref{eq:shiftB}) that we must sum over all topologically non-trivial
$\Sigma$-configurations in the path integral.

To implement such a gauge, we would naively introduce a $\delta$-functional
in Fourier representation by adding a term
\beq
-ib^a \bar{\chi}\biggl( \Sigma^\dagger T_a \Sigma \gamma_0 P_+ +
T_a \gamma_0 P_- \biggr) \chi
\label{eq:naive}
\eeq
(and an associated Faddeev-Popov determinant)
to the Lagrangian ${\cal L}'$. Such a gauge choice would precisely seem to
remove the
fermionic components of the shifted charge densities, as follows from
${\cal L}'$ if we take functional derivatives w.r.t. the vector
source $V_0$. However, in
order to obtain a well-defined BRST--gauge-fixed functional integral
one has to carefully take into account the regularized fermionic
measure. The ghost term corresponding to this gauge is traditionally
derived from an infinitesimal gauge variation; the additional coupling
to the fields $b^a (x)$, however, not only modify the Lagrangian but,
for consistency, also  the fermionic regulator. The whole gauge-fixing
procedure is then rather non-trivial (and analogous to the one encountered
in refs. \cite{us,us1}), but can in fact be solved by choosing the following
Lagrangian for the gauge-fixed functional integral:
\bea
{\cal L}'' &=& {\cal L}' (V_0 + b, \Sigma ) - b^a \frac{\delta}
{\delta V_0^a}
\biggl( {\cal L}_J (V_0,\Sigma ) + {\cal L}_{WZ} (V_0,\Sigma )\biggr)
\cr
&& - \bar{c}^a \nabla_b \frac{\delta}{\delta V_0^a} \biggl( {\cal L}_J
(V_0,\Sigma ) + {\cal L}_{WZ} (V_0,\Sigma )\biggr) c^b~.
\label{eq:Lgf}
\eea
Here, ${\cal L}'$ is the chirally rotated QCD Lagrangian including the
contributions from the chiral Jacobian as explained in
(\ref{eq:ZnewL}) and $\nabla_b$ denotes the Lie-derivative with
respect to $\Sigma$. The
explicit expressions turn out to be quite lengthy, but can be read off from
eqs. (16) and (17), so there is no need to display them here.
Note that the term {\em linear} in $b^a (x)$ is
precisely the naive gauge-fixing term (\ref{eq:naive}). The higher
order terms resulting from an expansion of ${\cal L}' (V_0 + b,\Sigma
)$ with respect to $b = b^a T_a$ guarantee that we have
precisely the gauge we want (the {\em physical} $J^a_0$ expressed entirely
in terms of bosonic fields only), once the non-invariance of the
fermionic measure has been taken into account.

After having fixed the gauge a global BRST-symmetry is left. The
corresponding transformations are
\bea
\delta \chi_L &=& ic^a T_a \chi_L \cr
\delta \bar{\chi}_L &=& -i\bar{\chi}_L c^a T_a \cr
\delta \Sigma &=& -i\Sigma c^a T_a \cr
\delta c^a &=& -\frac{1}{2} f^a_{bd} c^b c^d \cr
\delta \bar{c}^a &=& b^a \cr
\delta b^a &=& 0 ~.
\eea
Again, in order to see that this symmetry is realized in the path integral,
it is crucial to take into account the non-trivial Jacobian from the
regularized fermionic measure.

With the above gauge choice, baryon
density and in particular baryon number are carried by the
$\Sigma$-fields. This can be read off from the coupling to an external
singlet vector source. The fact that baryon number
is now carried by the $\Sigma$-field is enough to show that these
configurations behave as fermions. One can either follow the proof of
Witten \cite{Witten}, or use the following heuristic argument:
the statistics of a state of $N_cB_q$ quarks can be tested by
comparing the original S-matrix for
QCD with the one where in the final state the fermion fields are
replaced by their negative. This can be achieved by $e.g.$ an
adiabatic spatial rotation by an angle $2\pi$ or, better for our purposes, by
an adiabatic abelian phase rotation by an angle $\pi$. In the original
formulation of QCD such a transformation is given by
\beq
q(x) \to e^{i\pi t/T}q(x)~~,~~\bar{q} \to \bar{q} e^{-i\pi t/T}
\eeq
with $T$ being the time interval under consideration. This
transformation has only one effect: it modifies the zeroth component
of an external abelian vector source; the net effect is the addition
of a term $i\pi N_c B_q$ to the (euclidean) QCD-action with $N_c B_q$
the quark number. In the functional integral representing the
(euclidean) S-matrix this yields an additional factor $e^{i\pi N_c
B_q} = (-1)^{N_c B_q} $ as required for fermions. In our gauge fixed
version the quark number is replaced by the topological charge and
therefore the additional factor reads $ (-1)^{Q_{top}}$ with $Q_{top}$
the topological charge of the $\Sigma$-field.

So far our discussion has been restricted to the case of only one
light flavour. As argued in the introduction, the distinction between
``constituent'' and ``current'' in the case of the heavier quarks should be
fairly insignificant, and there should hence be no need to understand
such heavy constituent quarks by means of collective field
excitations.\footnote{Although the same scheme for extracting coloured
collective fields can be carried through for heavy quarks, the final
gauge-fixed picture in the analogue of the gauge (\ref{eq:Lgf}) would
presumably correspond to the constituent quark being described in
terms of a massive
$\Sigma$-excitation, with the mass scale given by the heavy Lagrangian
quark mass. Describing certain aspects of quark dynamics in terms of such
field configurations does not appear to be advantageous.} But the case of
several light quark flavours requires a separate discussion.
A natural choice \cite{Kaplan} of collective field
$\Sigma(x)$ in this case is to take it to be a group element of
both colour and flavour, namely (in our case)
$U(N_f\times N_c)$. Modifications of
the discussion above are then rather minor, and we comment here only on the
most important differences.

It is clear that
formally all expressions remain the same except that the trace is now
over colour {\em and} flavour. The essential point is that the
coefficient of the Wess-Zumino term is the same as in the one-flavour
case because $\Sigma$ is neither flavour- nor colour-blind and
therefore neither the colour nor the flavour trace yields an overall
factor $N_c$ or $N_f$. Therefore, the configurations with topological
charge one carry baryon number $1/N_c$ and behave as fermions, and
in this respect they have the quantum numbers of quarks.

The reader may well wonder what would happen if in the case of several
light quark flavours we instead chose to introduce collective fields
describing only the subgroup of $U(N_c)$, instead
of the bigger group discussed above. The most essential differences
can again be derived from the Wess-Zumino term. Now, it acquires a
factor $N_f$ because the field $\Sigma$ is flavour-blind. As a consequence
the baryon number associated with unit topological charge would be
$N_f/N_c$! Furthermore, the corresponding $\Sigma$-configurations
would behave as fermions only if the number of flavours $N_f$ is odd. From
this we see that to describe the correct quantum numbers of quarks requires
configurations with fractional topological charge $1/N_f$.
A priori there is nothing wrong with such configurations, except that
now we have to modify the $\Sigma$-integration so that we sum over
all topological sectors carrying integer multiples of topological
charge $1/N_f$.

The collective field formalism itself offers
no advice on how to choose new degrees of freedom in the fundamental
Lagrangian, and the above choice is as valid for a rewriting of QCD as one
based on fields that are elements of $U(N_f\times N_c)$. In fact, we are
obviously free to introduce collective fields transforming under
arbitrary groups, and are not at all restricted to some (approximate)
global symmetries.
It is only reasoning based on physical pictures
which favours certain choices. For example, in the case of $U(N_f\times
N_c)$ we introduce
collective field degrees of freedom for each flavour separately, in
contrast to the case of just $U(N_c)$. In
general, the only criterion for the choice of collective fields as
well as the choice of a gauge is efficiency; by this we mean that the
representation of the theory coincides as much as possible with the
physically relevant degrees of freedom. Such a criterion naturally leads
one, in the case of QCD flavourdynamics, to the introduction of collective
``mesonic'' fields as representations of $SU(N_f)\times SU(N_f)/SU(N_f)$
\cite{us}. For the present chiral-colour case this is not nearly as
obvious since neither current quarks nor constituent quarks appear
as asymptotic states, and a $q-\bar{q}$ state of $\Pi^a$ quantum numbers
would presumably not even be stable \cite{Int}.
Efficiency in this case means an accurate
description of physics by the most economical set of degrees of freedom
possible in the given context. A
description of strong interaction physics in terms of qualitons
interacting via gluon exchange may or may not stand up to this
criterion. In the four-dimensional case this still has to be shown.

Our derivation of the effective Lagrangian for QCD described in terms of
new coloured fields is, if all terms are kept in the $1/\Lambda$-expansion,
exact. With or without gauge fixing, we always regain cut-off QCD in its
original formulation if we integrate out the collective fields. Interpreted
as an effective theory, one must eventually face the problem that all
scales in the Lagrangian (for zero current quark masses) are set by the
ultraviolet cut-off $\Lambda$. In the renormalized theory this should be
replaced by a number of the order of the QCD $\Lambda$-parameter, in the
given scheme. The precise way such a renormalization must be carried out
is not known, and this is clearly one of the central problems with
the otherwise rigorous procedure described here. In the end, the
effective couplings to the collective fields should be entirely determined
by the strong interactions themselves.
Truncating the effective Lagrangian, and replacing all factors of the
ultraviolet cut-off $\Lambda$ by a parameter motivated by physical
intuition, may involve uncontrollable errors. So care should
be exercised when trying to extract approximate solutions from
the effective theory we have derived here. But it seems worthwhile to try
to derive a rough bound on, e.g., soliton masses in the present effective
theory, and to investigate the quantum stability of such solitons in general.

We finally wish to comment on an alternative route to understanding
the constituent quark problem which has recently
been suggested in ref. \cite{us}. There, the constituent quarks are viewed
as the {\em flavour}-chirally rotated fermionic remnants of the original
current quarks in the phase where chiral symmetry is spontaneously
broken, and the flavoured collective fields are identified with the
corresponding (pseudo-)Goldstone bosons. It is to a large degree
coincidental that
the procedure for extracting collective fields from the QCD Lagrangian
in both the colour-singlet \cite{us} and the colour-adjoint case
(the one considered here) has so many facets in common,
and the interpretation is
indeed entirely different. Thus, in ref. \cite{us} it is suggested that
the remnants of the quark
fields after extracting the pseudo-Goldstone bosons through flavour-chiral
collective-field phases may be directly identified with the constituent
quarks. In contrast, the idea of Kaplan involves an identification of
non-trivial topological excitations in bosonic variables with constituent
fermionic degrees of freedom. The effective cut-off Lagrangian we have
derived directly from QCD in this paper offers a picture very close to
this. The chirally rotated quark fields do, in the gauge presented
in eq. (\ref{eq:Lgf}), not give any contribution to the baryon number.
These rotated quark fields obviously still play a large part in
QCD dynamics, and in the gauge (\ref{eq:Lgf}) it is essentially
only their baryon
number density which is forced to vanish by the effective interactions.
For almost any other gauge choice we would, in order to be able to
compare directly with the scenario of Kaplan \cite{Kaplan}, have to
integrate out the remnant fields $\bar{\chi},\chi$ of the path integral.

These two alternative pictures of constituent quarks are clearly not
mutually exclusive, just as descriptions of baryons in terms of quark
bound states or Skyrmions are not. One may even conceive of intermediate
pictures, ``hybrid coloured bags'', in which one chooses to view
constituent quarks as bags surrounding current quarks, dressed up with
topological excitations of the $\Sigma$ field at intermediate distances.
Such a picture requires in the present framework the consideration of
a smooth gauge condition
that interpolates between the baryon number being carried entirely by the
$\bar{\chi},\chi$ fields at very short distances, to being carried by
the $\Sigma$
field at larger distances. One may even impose a bag wall by a step-function
gauge-fixing condition at the bag surface, analogous to the
Cheshire Cat gauges discussed in connection with ref. \cite{us1}. Then
baryon number would always add up to,
for $N_c = 3$, the appropriate 1/3 (as discussed

for the analogous flavoured case in ref. \cite{Rho}), despite being
carried by entirely different field configurations at short to intermediate
distances. The idea of even more exotic pictures can also be
entertained within the gauge-symmetric effective Lagrangian framework. All
these pictures should, if the gauges are chosen correctly, be physically
equivalent.

\vspace{0.5cm}

\noindent
{\sc Acknowledgment:} We would like to thank D. Kaplan, R.D. Ball and
Y. Frishman for discussions.

\newpage

\end{document}